\documentclass[twocolumn,showpacs,preprintnumbers,prb,amsmath,amssymb]{revtex4}

\usepackage{graphicx}
\usepackage{dcolumn}
\usepackage{bm}

\begin{document}

\title{Extracting the Pair Distribution Function of Liquids and Liquid-Vapor Surfaces by Grazing Incidence X-ray Diffraction Mode}

\author{David Vaknin, Wei Bu, and Alex Travesset}
\affiliation{Ames Laboratory, and Department of Physics and Astronomy, Iowa
State University, Ames, Iowa 50011, USA}
\date{\today}

\begin{abstract}
We show that the structure factor $S(q)$ of water can be obtained from X-ray synchrotron experiments at grazing angle of incidence (in reflection mode) by using a liquid surface diffractometer.  The corrections used to obtain $S(q)$ self-consistently are described.  Applying these corrections to scans at different incident beam angles (above the critical angle) collapses the measured intensities into a single master curve, without fitting parameters, which within a scale factor yields $S(q)$.  Performing the measurements below the critical angle for total reflectivity yields the structure factor of the top most layers of the water/vapor interface. Our results indicate water restructuring at the vapor/water interface.  We also introduce a new approach to extract $g(r)$, the pair distribution function (PDF), by expressing the PDF as a linear sum of Error functions whose parameters are refined by applying a non-linear least square fit method.  This approach enables a straightforward determination of the inherent uncertainties in the PDF.    Implications of our results to previously measured and theoretical predictions of the PDF are also discussed.
\end{abstract}

\maketitle
\section{Introduction}
The pursuit after the structure and properties of water is obviously driven by its importance as the medium that supports life on Earth. Structural studies were initiated from the outset, after the discovery of X-rays and neutrons\cite{Hansen2003}, and have continued to these days, as technological advances, such as improved X-ray or neutron sources and computational capabilities have enabled more precise and refined insights\cite{Head-Gordon2002,Abbamonte2004}. High energy X-ray diffraction techniques for studying liquids and glasses\cite{Poulsen1995,Mei2007} have also advanced the field\cite{Hart2005}.   The majority of previous and current X-ray studies of bulk water have been performed in {\it transmission} mode, where an X-ray beam travels through the windows of a container before and after scattering from the sample or directly scattering through a flowing cylindrical jet of the liquid\cite{Katzoff1934,Morgan1938,Narten1971,Nishikawa1980,Gorbaty1985,Soper1986,Soper1997,Hura2000,Sorenson2000}.
Narten and Levy and co-workers introduced a Bragg-Brentano type diffractometer to collect the diffraction pattern from a horizontal liquid surface in reflection mode that eliminates sample holder absorption and scattering\cite{Levy1966}. Herein, we extend on the reflection mode by using the grazing angle X-ray diffraction (GIXD) mode.  To conduct experiments in GIXD mode, we take advantage of the liquid surface (horizontal) diffractometer, that was first introduced by Als-Nielsen and Pershan\cite{Als-Nielsen1983}.

The quantity of interest in structural studies is the liquid structure factor (or the absolute scattering
cross-section\cite{Page1971}) $S(q)$, from which the pair distribution function (PDF) $g(r)$ is extracted.  Experimentally, however, the structure factor is only obtained after accounting for multiple corrections and assumptions, which may introduce systematic errors into the final results
\cite{Karnicky1976,Head-Gordon2002}.  The possibility of providing a significant simplification in the measurement of liquid structure-factors was one of the motivations to pursue the structural studies described in the present study.  Compared with transmission mode, our experiments in reflection mode do not require corrections such as the ones related to the presence of the container or $1/r^2$ geometric considerations and corrections due to absorption and background scattering become substantially simpler\cite{Levy1966}.  In this study, we determine the structure factor of water and compare our results with previous experiments. More importantly, as the real part of the index of refraction for X-rays is smaller than unity, the x-ray beam undergoes total external reflection below a critical angle $\alpha_c$, with a finite penetration depth. Thus, scans above the critical angle predominantly provide the bulk water structure factor, while scans below the critical angle probe the structure of the top most layers of the liquid surface, providing direct information on the restructuring at the water/vapor interface.

\section{Determination of $S(q)$ in reflection mode}
\subsection{Reflection mode setup}
\begin{figure}[htl]
\includegraphics[width=2.6 in]{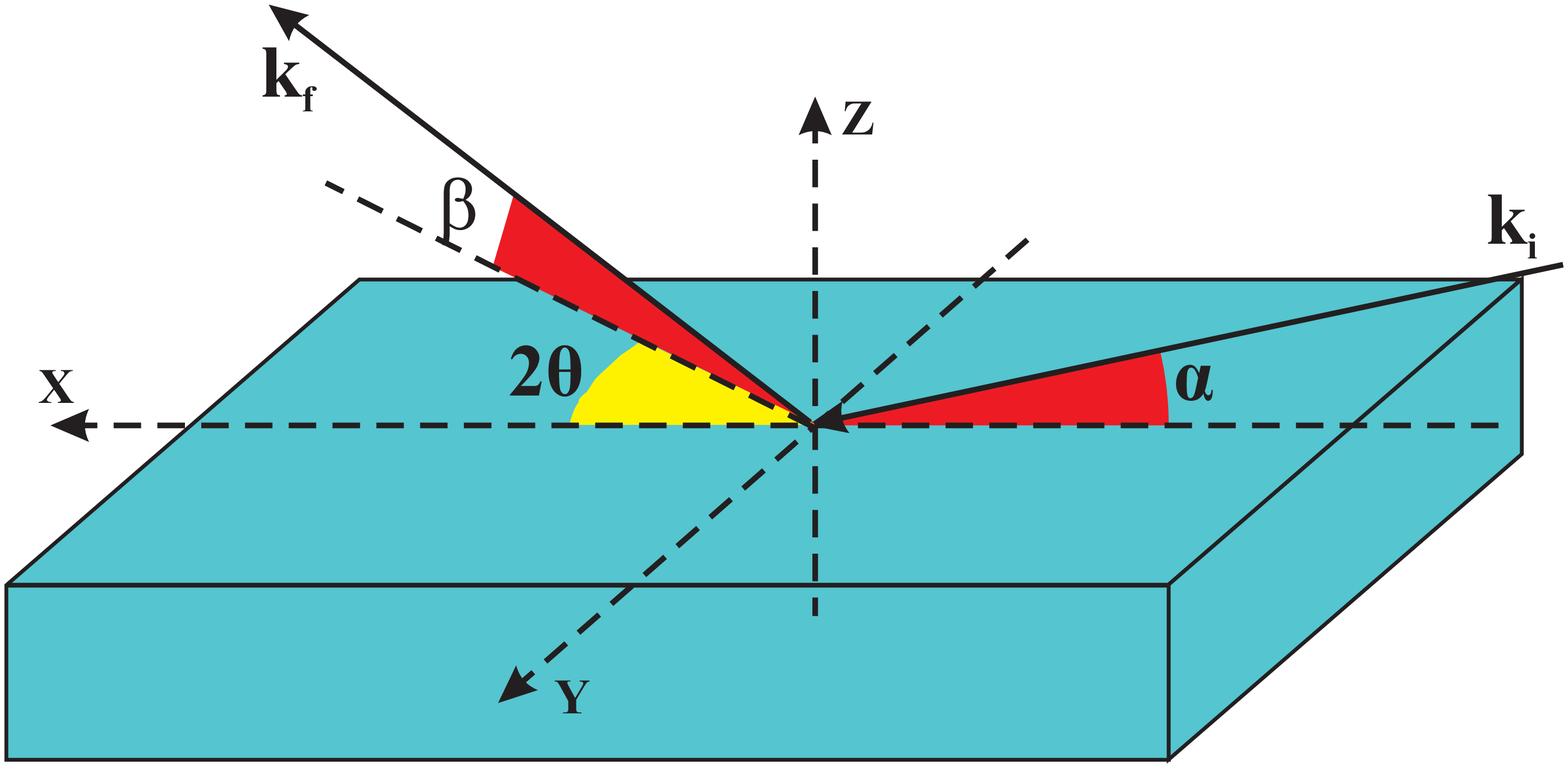}
\caption{\label{setup}Setup for measuring bulk structure factor of a liquid in reflection mode. The incident beam with a wave-vector {\bf k$_i$} hits the surface at a fixed angle $\alpha$ with respect to the liquid surface.  For bulk measurements $\alpha$ is kept above the critical-angle for total external reflection. The scattered beam is collected at an angle $\beta$ with respect to the surface and at an angle $2\theta$ with respect to the X-axis in the XY-plane. }
\end{figure}

Figure\ \ref{setup} shows the setup used to measure the structure factor in reflection GIXD mode from a liquid.  The incoming beam (propagating along the X-axis) hits the flat liquid surface at an angle of incidence $\alpha$ with respect to the surface, and the scattered beam is detected at an angle $\beta$ with respect to the surface and an azimuthal angle $2\theta$ measured from the X-axis.  The scattering vector is given by
\begin{equation}
\textbf{q} = k_0\left(\cos{\beta}\cos{2\theta}-\cos{\alpha},-\cos{\beta}\sin{2\theta},\sin\alpha+\sin{\beta}\right).
\end{equation}
where $k_0 = 2\pi/\lambda$, and $\lambda$ is the X-ray wave-length. Because of the isotropic nature of the system, scattering scans are presented as a function of the modulus $q\equiv|\textbf{q}|$
\begin{equation}
q = k_0\sqrt{2+2\sin\alpha\sin\beta-2\cos\alpha\cos\beta\cos2\theta},
\end{equation}
which for $\alpha=\beta=0$ gives the known expression $q=2k_0\sin\theta$.

\begin{figure}[thl]
\includegraphics[width=2.6 in]{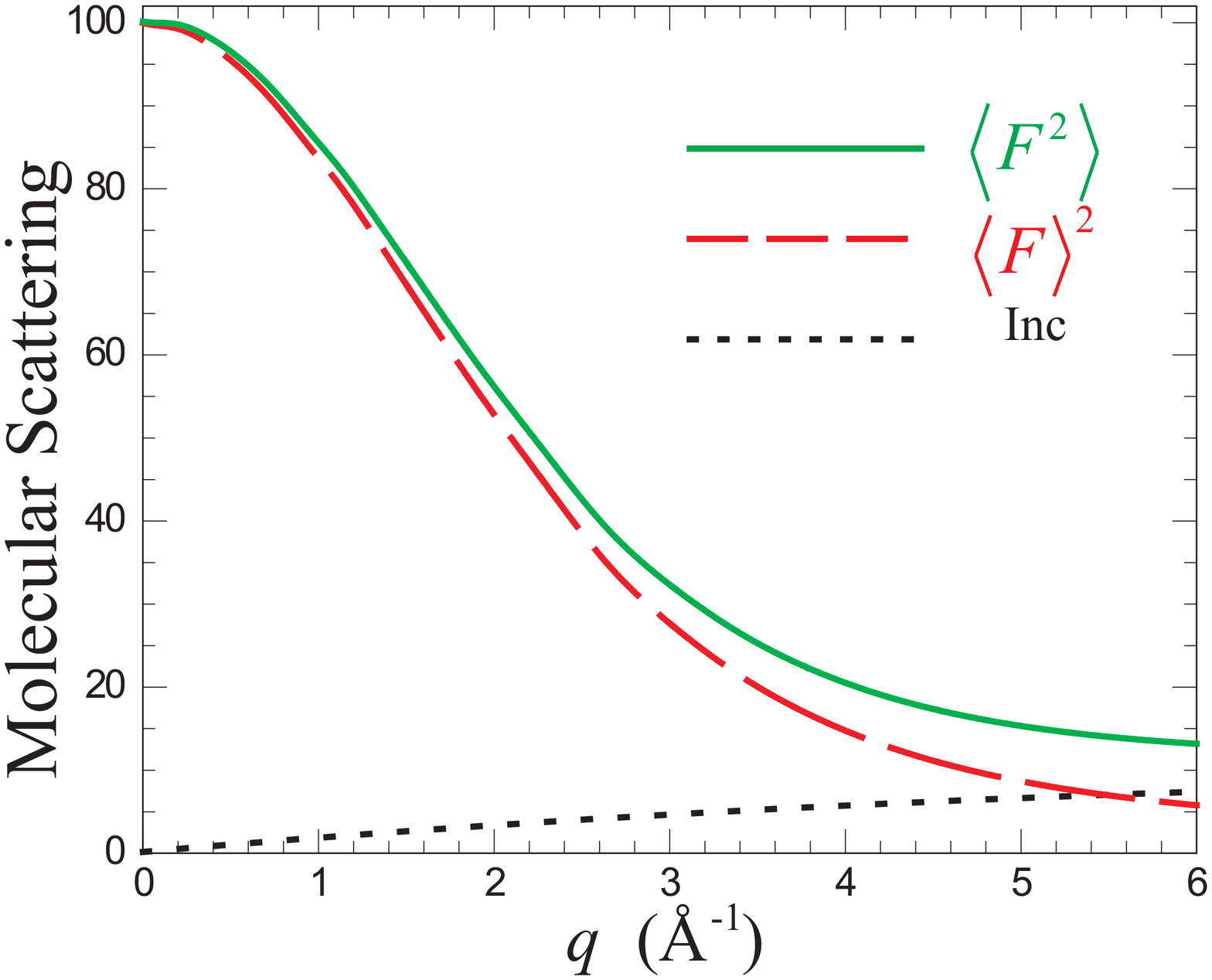}
\caption{\label{sfactor} The coherent form factor $\langle F(q)\rangle^2$ (dashed line) and the total incoherent Compton scattering (dotted line) as given in Ref.\ \onlinecite{Wang1994}.  The solid line shows the sum of the incoherent and coherent scattering $\langle F(q)^2\rangle$.}
\end{figure}

We use standard calculations for the liquid structure factor\cite{Guinier1994}.  In reflection mode, the absolute scattering cross section includes a contribution due to scattering from capillary waves\cite{Sinha1988}, however this scattering is negligibly small compared to bulk scattering above the critical angle, and can be neglected. The relation between the structure factor and the PDF for bulk water (above the critical angle) is
\begin{equation}
S(q) =  \langle F^2\rangle  +  \langle F\rangle ^2 \int\limits_0^\infty  {4\pi \rho r^2 } \left[g(r) - 1\right]\frac{{\sin (qr)}}{{qr}}dr
\label{SQ}
\end{equation}
$\rho$ is the number density of the liquid.  $\langle F(q)\rangle$ is the average form factor of an H$_2$O molecule (see Fig.\ \ref{sfactor}).  The $\langle F(q)^2\rangle$ is the sum of the coherent and  incoherent scattering from each molecule,
\begin{equation}
\langle F^2\rangle = \langle F\rangle^2+I_{CS}(q)
\end{equation}
where $I_{CS}$ is the inelastic Compton scattering.  For the average form factor and the incoherent Compton scattering we use the calculated result given in Ref.\ \onlinecite{Wang1994} which does not significantly differ from that given in Ref.\ \onlinecite{ITOC-III}.

The raw measured intensity $I(q)$ depends on variables such as the incident angle of the X-ray beam and on the X-ray wavelength. Its relation to the structure factor $S(q)$ is given by
\begin{equation}
S(q) = C\frac{I(q)-I_B(q)}{I_0V_{eff}(q;2\theta,\alpha,\beta)P(q;2\theta,\alpha,\beta)}
\label{I_to_SQ}
\end{equation}
where $C$ is a scale factor (determined by our analysis described below) $I_B$ is the background intensity,  measured by lowering the surface below the incident beam and otherwise conducting the same scan as with the sample.  $I_0$ is the intensity of the incident beam on the sample which is monitored immediately before the sample to account for fluctuations in the source, and due to the configuration of the instrument.  $V_{eff}(q;2\theta,\alpha,\beta)$ is the effective volume of scattering, and $P(q;2\theta,\alpha,\beta)$ is the polarization factor.  In reflection mode, the angles  $\alpha$ and $\beta$ are generally very small ($\lesssim 3$ degrees). Therefore, to a good approximation, the polarization factor is given by,
\begin{eqnarray}
P(q;2\theta,\alpha,\beta) &\approx (1+\cos^2{2\theta})/2  & \textrm{--  unpolarized} \nonumber\\
    &\approx \cos^2{2\theta}   & \textrm{--  polarized},
\label{polarization}
\end{eqnarray}
The evaluation of the effective volume of scattering is described in detail in Appendix A.

\section{\label{sec-method} A new method of extracting $g(r)$ from $S(q)$ for bulk water}
The aforementioned corrections, to within a scale factor, yield $S(q)$ from which the PDF has to be extracted by solving Eq.~ (\ref{SQ}).  Traditionally, the method to accomplish this is by inverse integration of Eq.\ (\ref{SQ}), which requires extrapolation to $q-$values where $S(q)$ has not been measured. This extrapolation is not free from technical problems, as described, for example, in Ref.\ \onlinecite{Narten1971} Eq.\ (9) and (10).

Herein, we introduce a new procedure for obtaining the PDF (for bulk water) that overcomes the problems inherent to the inverse integration. The method works as follows; we construct a model function $g(r)$ that is generated by parameterized Error functions
\begin{equation}\label{erf}
g(r) =
\frac{1}{2}\sum_{i=1}^{N+1}(G_{i+1}-G_i)\textrm{erf}\left(\frac{r-r_i}{\sigma_i}\right)+\frac{1}{2}G_{N+2}.
\end{equation}
The conditions, $g(r)= 0$ at the origin and $g(r)= 1$ for large $r$, imply $G_1=0$ and  $G_{N+2}=1$, respectively.  The method starts with one Error function ($N=0$) the parameters of which are refined by a non-linear square fit (NLSF) method that minimizes a global quality factor $\chi^2$ to obtain the best fit to the numerically calculated $S(q)$ Eq.\ (\ref{SQ}).  An Error function is added iteratively one by one, incrementally increasing $N+1$ (and the number of parameters).  Each additional Error function adds three more parameters namely, $G_i$, $r_i$ and $\sigma_i$.  The number of Error functions $N$  used at the end of the process is the minimum number necessary to fit the data such that the addition of another Error function (with its corresponding parameters) does not improve the quality factor of the fit $\chi^2$.  The scale factor $C$ in Eq.\ (\ref{I_to_SQ}) (dependent on incident beam intensity) is also a free parameter that is refined in this process, it is dominated by the number density of water molecules and the molecular form factor (elastic and inelastic).
Alternative techniques to directly calculate the PDF from the experimentally determined structure factor have been introduced in the past\cite{Soper1996}.  For instance, in the empirical potential structure refinement technique (EPSR), the parameters of interatomic potential energy function are refined to produce the best fit between the simulated and measured structure factor\cite{Soper1996,Hura2003}.  Our approach differs in that it does not rely on any theoretical assumptions.

This process allows to determine uncertainties of the free  parameters $G_i$, $r_i$ and $\sigma_i$ which in turn yield the uncertainties in $g(r)$.  The spread in the values of $G_i$, $r_i$ and $\sigma_i$ reflects the uncertainties associated in extracting $g(r)$ from a the error-bars of each point in $S(q)$ and the finite $q$-range  in $S(q)$.  In other words, the analysis of the experimental result does not yield a single PDF, but a spread of functions that decode quantitatively the inherent uncertainties of the experimental results.

\section{Experimental Details}
The X-ray scattering experiments were conducted on the Ames Laboratory Liquid Surface Diffractometer at the 6ID-B beamline at the Advanced Photon Source at Argonne National Laboratory\cite{Vaknin2003,Vaknin2003a}. The highly monochromatic beam (16.2 keV with energy resolution, $\Delta E \sim 2\,\mathrm{eV}$), selected by an initial Si double crystal monochromator, is deflected onto the liquid surface at a specified angle of incidence by a secondary monochromator (Ge(220) single crystal), which is placed on the diffractometer\cite{Vaknin2003}.  The synchrotron X-ray beam is highly polarized ($\approx 98\%$) with the electric field parallel to the liquid surface, therefore, the polarization correction (Eq.\ (\ref{polarization})) is practically given by $P(q;2\theta,\alpha,\beta) =\cos^2(2\theta)$.

Ultrapure water (NANOpure, Barnstead; resistivity, 18.1 M$\Omega$cm) was used in the present study.  The water was contained in a Teflon trough, and a glass plate (area 10 x 5 cm$^2$) was placed in the trough to form a thin water film ($\approx$ 0.3 mm thick) to reduce the effect of mechanical agitations on the surface smoothness.   To test the quality of the surface we routinely checked that the reflection from the surface below the critical angle is nearly 100\%. The trough was encapsulated in an air-tight thermostated aluminum enclosure (T = 294 K), which was continuously purged with a flow of helium gas (bubbled through water) during the course of the experiment to lower background scattering from air.  The height of the water surface with respect to the beam center was determined by a reflection from the surface to better than $\pm 0.005$ mm.
\section{Results and Analysis}
\subsection{Bulk Water}
\begin{figure}[t!p]
\includegraphics[width=2.4 in]{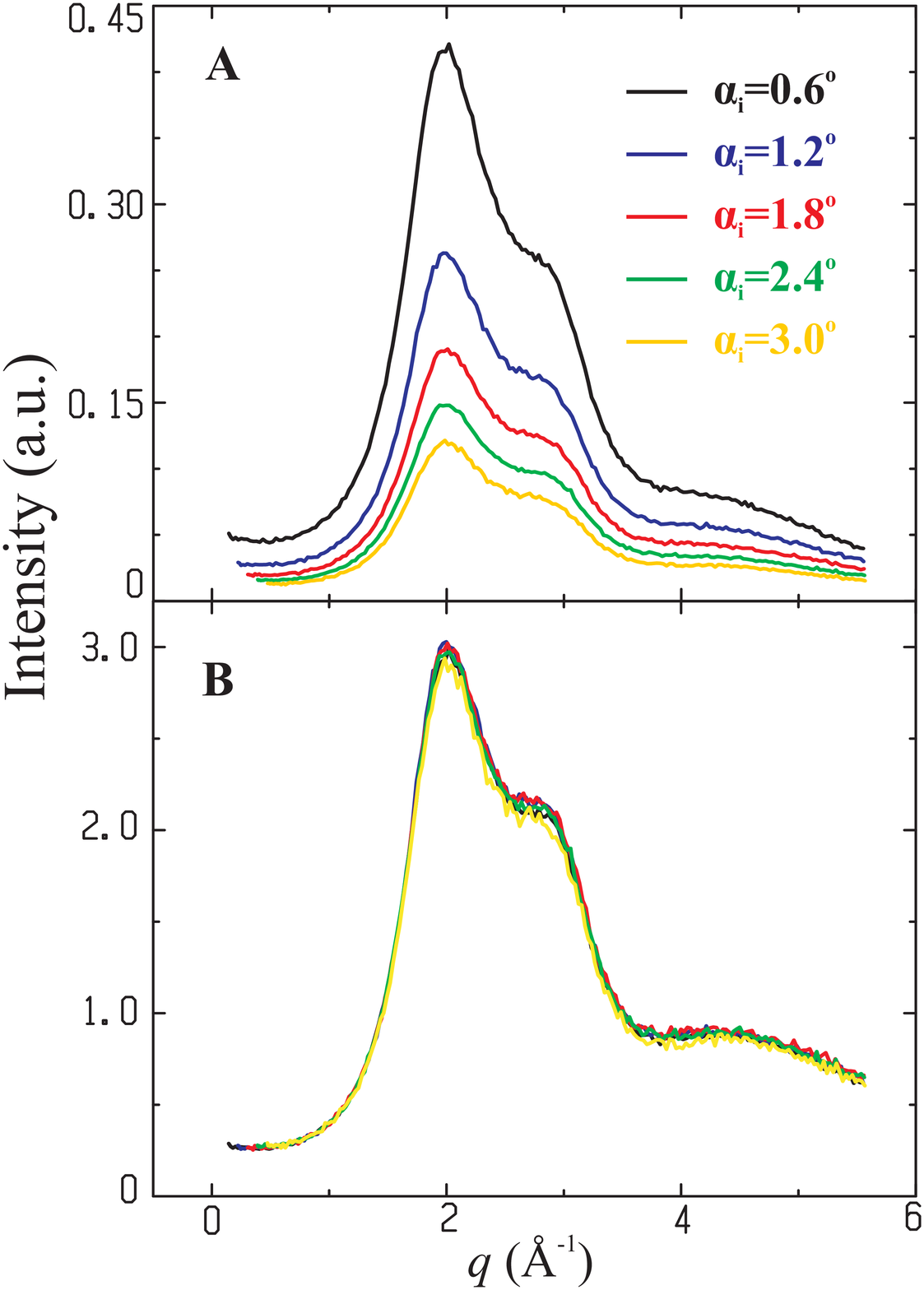}
\caption{\label{SF} (A) Scattered intensities versus momentum transfer $q$ for various incident-beam angles $\alpha$ at $T=294$ K. The background at all $q$ values is at the $10^{-4}$ level, in the same units as shown in the figure. (B) Same data after normalization by the effective volume of scattering $V_{eff}$.  All the data at different incident angles collapse to a single master-curve without any fitting parameters.  The data are also corrected by the polarization factor.  This is $S(q)$ up to a scale factor.}
\end{figure}

Figure\ \ref{SF}(A) shows the measured raw intensity versus $q$ data at several incident beam angles $\alpha$ after normalization to the incident-beam monitor and after background subtraction. It is  evident that the raw data is dependent on the incident beam angle $\alpha$.  Normalization of the raw data by the effective volume of scattering $V_{eff}$ Eq.\ ~(\ref{Veff}) and by the polarization collapses all data sets at different angles $\alpha$ into a single master curve as shown in Figure~ \ref{SF}(B).  This collapse of the curves does not involve any adjustable parameters which provides a very stringent self-consistent test on the robustness of $S(q)$ (within a scale factor).

\begin{figure}[thl]
\includegraphics[width=2.4 in]{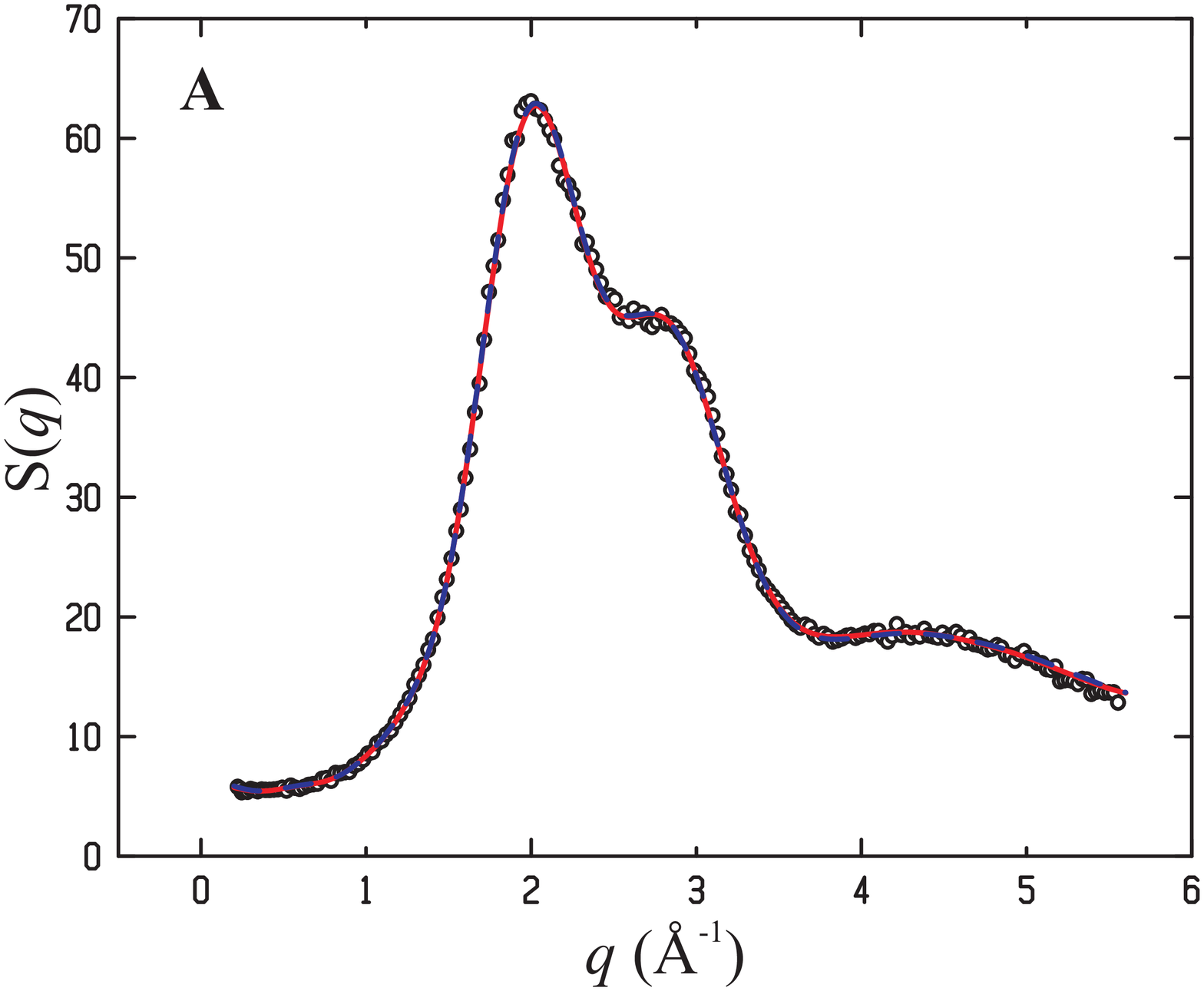}
\includegraphics[width=2.4 in]{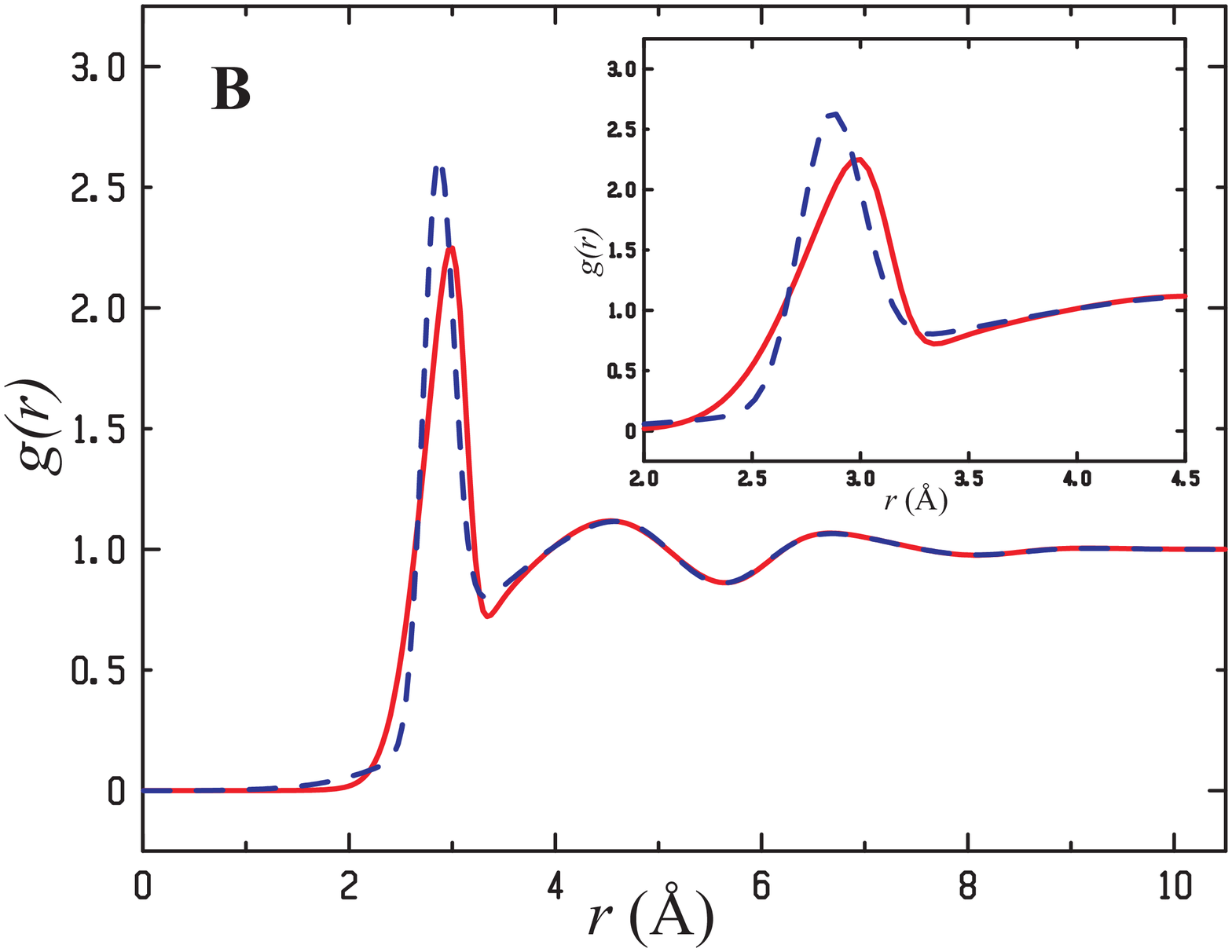}
\includegraphics[width=2.4 in]{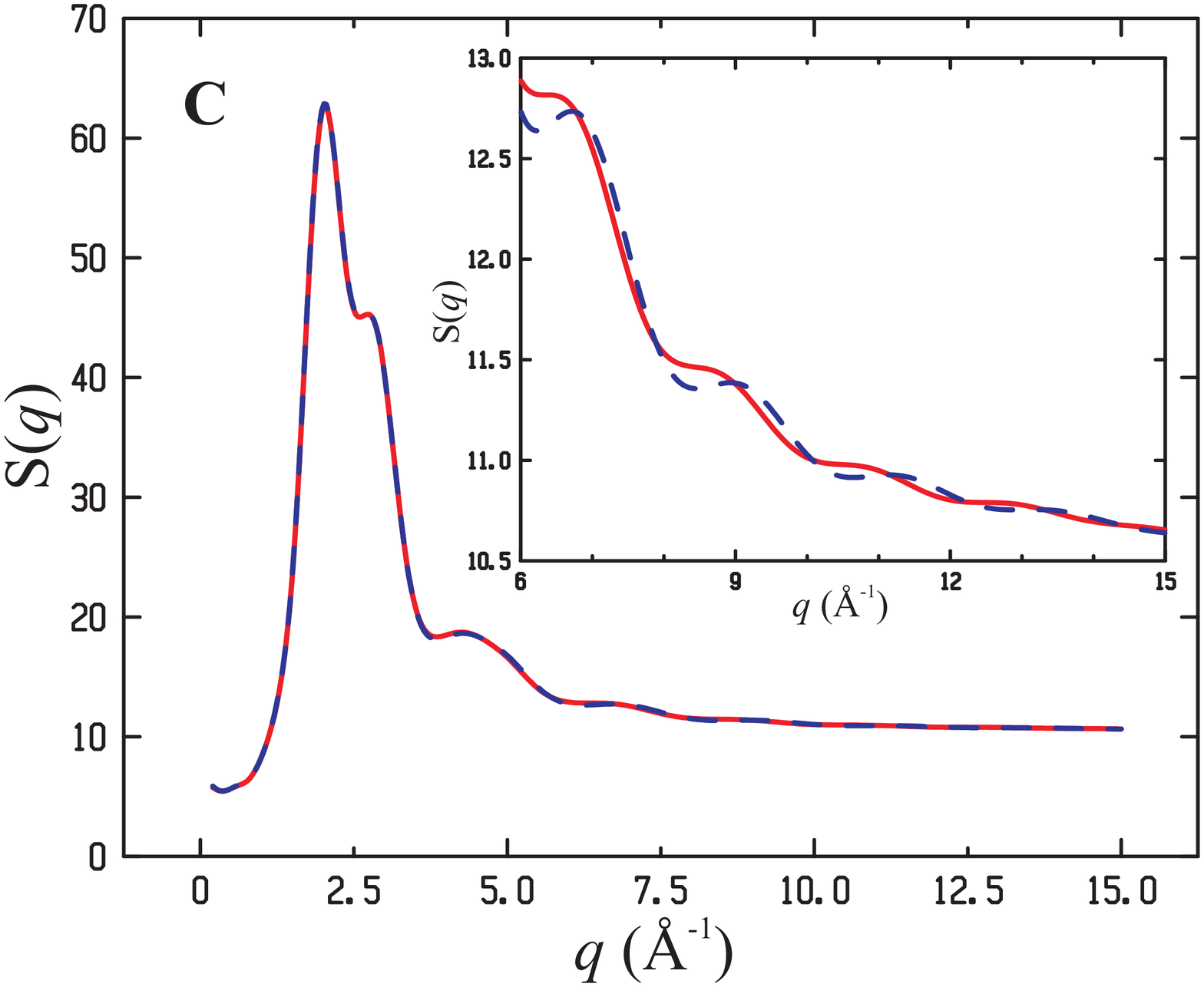}
\caption{\label{fit_gr} (A) $S(q)$ (at $T=294$ K) obtained after scaling the data shown in Fig.\ \ref{SF} (circles), the best fit (solid line) and a second fit (dashed line) with different PDF but within the uncertainty range.   (B) Two PDF's used to calculate the best fit the $S(q)$ shown in (A).  (C) $S(q)$ calculations extended to large $q$ values using the two PDFs shown in (B) showing the high $q$ values of $S(q)$ are almost identical.}
\end{figure}

To determine the PDF, we apply the NLSF method described in Section III to the master-curve shown in Fig.\ \ref{SF}, and in the process we also determine the scale factor $C$ (Eq.\ (\ref{I_to_SQ})) which yields the $S(q)$ shown in Fig.\ \ref{fit_gr}(A).  The scale factor determines, without any assumptions, the maximum value of $S(q)$ at $q \approx 2.0$ {\AA}$^{-1}$ is $63 \pm 1$ which compares well with previously reported values\cite{Narten1971,Sorenson2000,Hura2000}.  The solid line in Fig.\ \ref{fit_gr}(A) is obtained from the best-fit parameters listed in Table\ \ref{param} with their uncertainties.  The dashed line is obtained by small variation of parameters, within the determined errors.  In general we find that the most prominent uncertainty in the determination of the PDFs is confined to the region of the first peak ($\approx$ 2.8 {\AA}), which is mainly due to the finite $q-$range of the measurement.  Whereas the goodness of the fit is less sensitive to small variations in the height and the width of the first peak individually, their correlated value, associated with the number of nearest neighbors (NN) to a given water molecule is.  This number of NN is expressed by the sum rule relation
\begin{equation}
NN =4\pi\rho\int_0^{r_\text{min}}{g(r)r^2dr},
\label{NN}
\end{equation}
where $r_\text{min}$ is the location of the first minimum of the PDF.  Figure\ \ref{fit_gr}(B) shows the PDF producing the best fit to the data (solid line) and another fit within the allowed Errors (dashed lines) given in Table \ref{param}.  Despite the spread in the height and width of the first peak, all PDF gave within error the same number of nearest neighbors 4.7 $\pm 0.1$.  Figure\ \ref{fit_gr}(C) shows calculated $S(q)$ to large $q$-values (beyond our measurement range) using the two model PDFs shown in Fig.\ \ref{fit_gr}(B).  It should be noted that despite slight disagreements about the shape and position of the first peak, the present and the above mentioned studies all agree about the PDF at distances larger than $r_{min}$.
\begin{table}[htl]
\caption{\label{param} Parameters that generate the best-fit calculated
structure factor using Eq.\ (\ref{erf}.)}
\begin{ruledtabular}
\begin{tabular}{llll}
$i$ & $r_i$ ({\AA}) & $G_i$   &   $\sigma_i$ ({\AA})\\
\hline
1       &2.78   $\pm 0.02$     &0.00             &0.29  $\pm 0.16$\\
2       &3.13    $\pm0.15$     &3.01 $\pm 1.20$  &0.1    $\pm 0.04$\\
3       &3.60  $\pm0.02$       &0.49 $\pm 0.09$  &1.01  $\pm 0.33$\\
4       &5.10    $\pm 0.21$    &1.23 $\pm 0.01$  &0.61 $\pm 0.09$ \\
5       &6.06   $\pm0.02$      &0.79 $\pm 0.02$  &0.47   $\pm 0.03 $\\
6       &7.34  $\pm0.01$       &1.14 $\pm 0.01$  &1.39  $\pm 0.03$\\
7       &8.40    $\pm 0.01$    &0.90 $\pm 0.01$  &0.58 $\pm 0.05$    \\
8       &                      &1.00            &0.00\\
\end{tabular}
\end{ruledtabular}
\end{table}
\subsection{Restructured Water Surface}
\begin{figure}[!h]
\includegraphics[width=2.8 in]{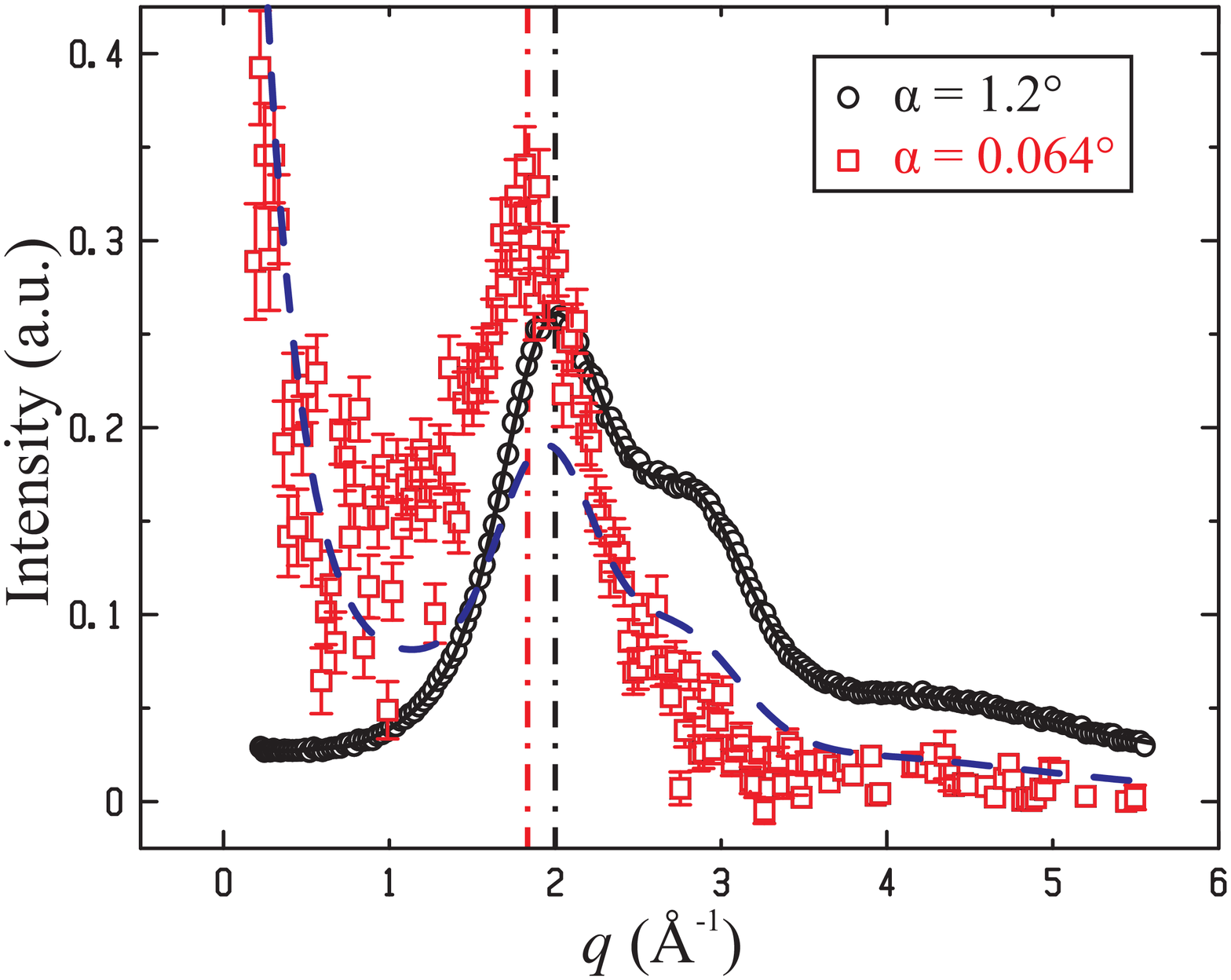}
\caption{\label{sur1} Raw GIXD data after background subtraction above and below the critical incident angle for total reflectivity as indicated. Solid lines are the best fit as discussed in the text. Vertical dashed-dotted lines indicate main peak positions of the bulk and surface structure factor.}
\label{surf_sf}
\end{figure}

Performing the GIXD scans below the critical angle provides a pattern that is highly surface sensitive due to the finite penetration depth of the evanescence wave (the penetration depth at $\alpha=0.064^{\circ}$ is $\approx 80$ {\AA}.), and the enhancement by multiple scattering, as predicted by the distorted wave Born approximation (DWBA) \cite{Vineyard1982,Kjaer1994,Vaknin2003}.  Figure\ \ref{surf_sf} shows two diffraction patterns above and below the critical angle for total reflection.   The scattering from the surface differs from that of the bulk (($\alpha>\alpha_{c}$)) in several respects.  First, as $q \rightarrow 0$ the intensity diverges due to diffuse scattering from surface capillary waves\cite{Sinha1988}.  Second, the main peak of bulk water structure factor, at $q_\textrm{max}\approx 2.0$ {\AA}$^{-1}$, is slightly shifted to smaller $q$ values suggestive of a larger intermolecular distances at the interface compared to bulk water.  This is in agreement with recent extended x-ray absorption fine structure spectroscopy (EXAFS) measurements of water microjets, that show the intermolecular O–O distance is 5.9\% larger than that of bulk water\cite{Wilson2002}.  Third, the shoulder ($q \sim 2.7$ {\AA}$^{-1}$) is less pronounced compared to that of bulk water.

In the following we attempt to modify the method described in Section\ \ref{sec-method} for a half-filled space with bulk water to examine whether geometrical effects (truncation of electron density at $z=0$) can give rise to the observed differences between the scattering from bulk and  surface.  In the absence of a depth-dependent PDF near the surface, we assume that the bulk $g(r)$, obtained in the present study, is valid everywhere including the surface. We argue that the scattering consists of two parts, as follows,
\begin{equation}
I(\alpha,\beta,2\theta)=I_{CW}(\alpha,\beta,2\theta)+I_{b}(\alpha,\beta,2\theta),
\label{eq_sur1}
\end{equation}
where $I_{CW}$ and $I_{b}$ represent capillary-wave diffuse scattering and {\it bulk} scattering contributions, respectively.  The two terms are derived in detail in Appendix\ \ref{App-2}. Numerically evaluating Eq.\ (\ref{eq_sur1}) and the complementary equations in Appendix\ \ref{App-2}\ with $a$ and $b$ as the sole parameters, and $g(r)$ as defined in Table\ \ref{param} (as shown in Fig.\ \ref{fit_gr}(B), we obtain a poor fit to the data (dashed-line in Fig. \ \ref{sur1}).  This shows that, Equation\ (\ref{eq_sur1}) although adequately describes the diffuse scattering it does not predict the observed shift of peak position at ($q \approx2.0$ {\AA}$^{-1}$) or the observed change of the feature at $q \approx 2.7$ {\AA}$^{-1}$.  This implies that the PDF of the top most layers is not the same as that  of the bulk PDF, and a more refined calculation considering an anisotropic $g(r_{||},z)$ needs to be used.

\section{Discussion and Summary}
\subsection{Comparison with previous determinations of the water PDF}
\begin{figure}[thl]
\includegraphics[width=2.4 in]{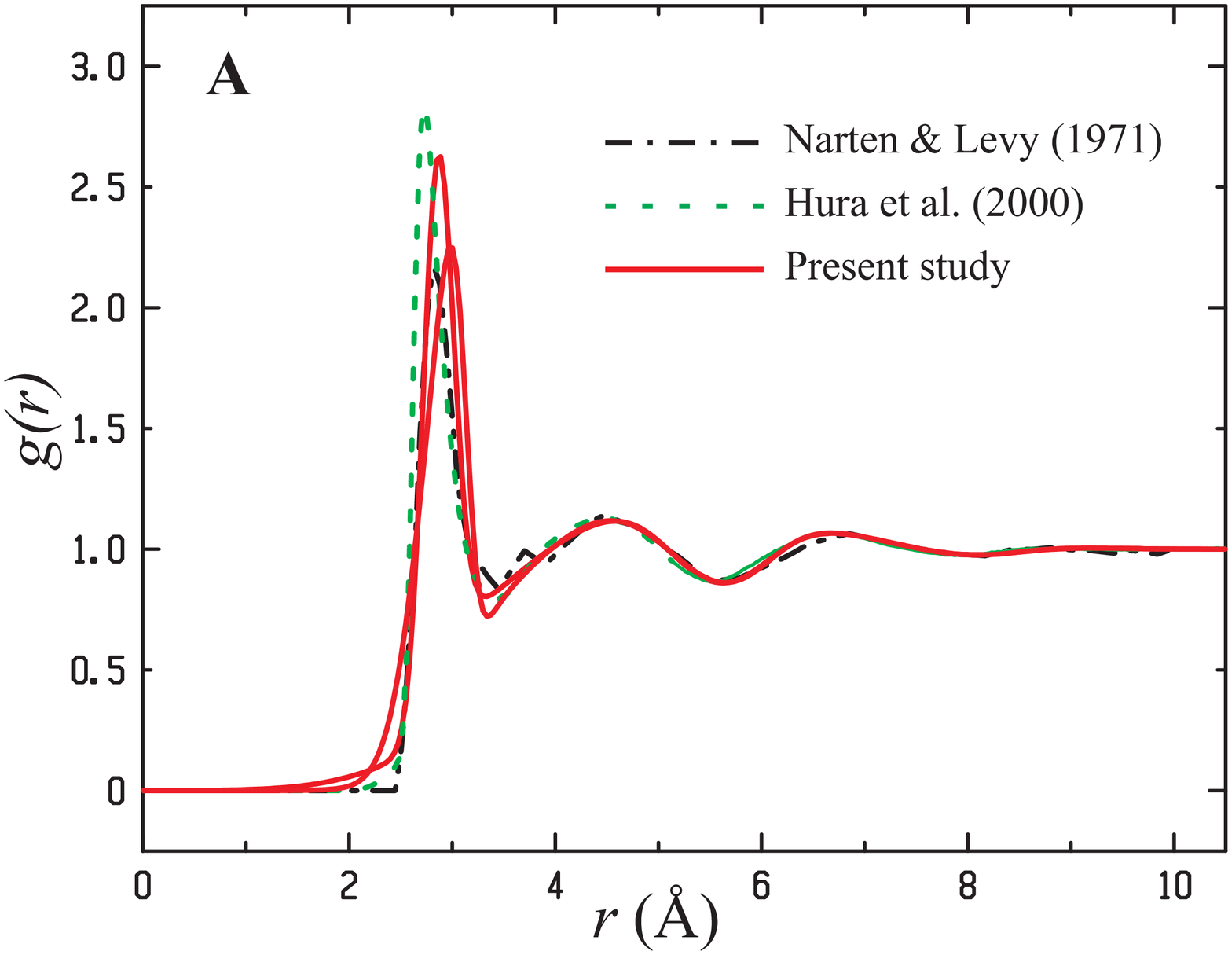}
\includegraphics[width=2.4 in]{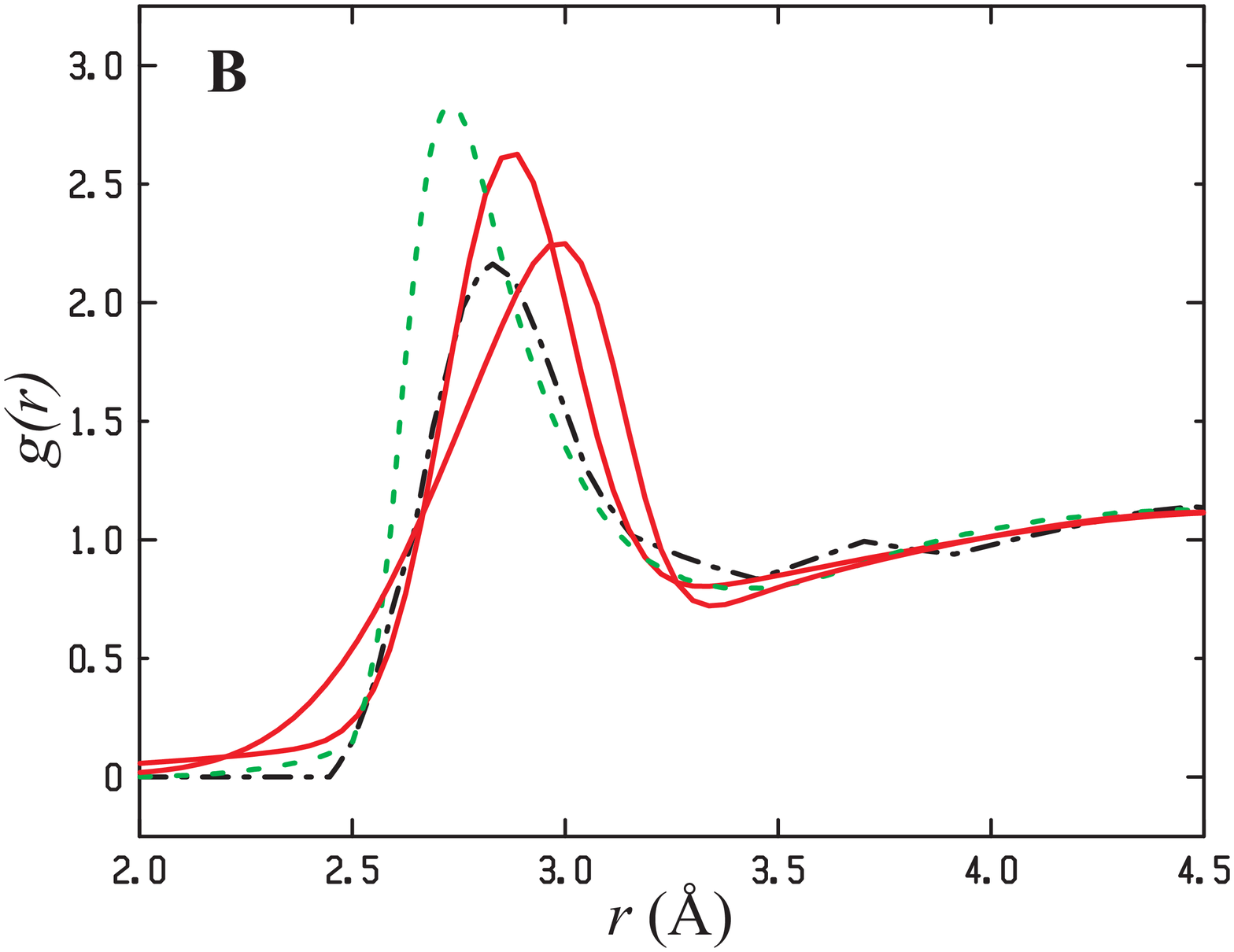}
\caption{\label{comparison}  A comparison of the PDF's from our study and previous studies as indicated.  The two PDFs of our study are the same as those shown in Fig.\ \ref{fit_gr}(B).}
\end{figure}

As discussed in the previous section, our determined PDF is not a single function but a spread of functions differing mainly in the height and width of the first peak but preserving the number of nearest neighbors (NN).  In Fig.\ \ref{comparison} we compare two of our PDF models with previous results\cite{Narten1971,Hura2000,Sorenson2000}.  Within error, both results are consistent with our determined PDFs. It may be argued that the relatively large dispersion on the peak height that follows from our results is a consequence of the relatively smaller range of $q$-values measured.  We have therefore applied the method to compute the PDF described in Section III to the data of Hura et al. Ref.\ \onlinecite{Hura2000}. The results, shown in Fig.\ \ref{Comp_hura} of Appendix \ref{App-3}, display a somewhat reduced but still quite significant dispersion.
 In particular, the number of NN  $4.7 \pm 0.1$ is the same as that obtained by Hura {\it et al.}  (NN = 4.7) but slightly different than that obtained by Narten and Levy (NN = 4.4)

Based on our method for extracting PDFs, we argue that it will be a daunting task to reduce the uncertainty in the first peak of the PDF as very precise measurements to large $q-$values, with accurate control over systematic and other errors is hard to achieve as the expected signal is extremely low. Figure\ \ref{fit_gr}(C) shows the minute differences between two possible $S(q)$'s at large $q$-values, which give rise to relatively large differences in the first peak of the PDF as shown in Fig.\ \ref{fit_gr}(B).
\subsection{Implications for theoretical models of water}
Accurate theoretical models of water are of fundamental importance for many problems in the physical properties of water, in particular in relation to biological processes. In Ref.\ \onlinecite{Sorenson2000}, the most popular water models were compared in great detail with the experimental results for the PDF to assess their validity.  The main conclusion drawn was that the TIP5P model\cite{Mahoney2000} provided the most accurate description of the experimental data.

There are some general implications that follow from our analysis in this regard. First, our analysis shows that the height and width of the first peak of the water PDF does not provide a very stringent test to validate theoretical models, as existing experimental results cannot accurately resolve these parameters, a point that was also noted by Jorgensen and Tirado-Rives\cite{Jorgensen2005}.
A more stringent test is to compare the PDF beyond the first minimum, as all recent experimental determinations, including ours, consistently yield almost indistinguishable PDFs. In this respect, the very popular TIP3P model\cite{Jorgensen1983} does not compare favorably with experimental data while the SPC/E \cite{Berendsen1987}, the other widely used model, compares well, and is more consistent with experimental results.  A more constraining test on the validity of theoretical models at distances lower than the first minimum in the PDF is provided by the number of NN, which our result places at 4.7 $\pm$ 0.1.  Both TIP5 and SPC/E are consistent with these numbers but TIP3P gives 5.1, significantly larger than the experimentally extracted value. Given its simplicity, it is remarkable how well SPC/E model matches the measured PDF. We conclude that whereas experimentally determined water PDFs provide valuable guides to test the validity of water models, there is still some uncertainty in the experimental region at distances smaller than the shell of NN i.e., the first minimum in $g(r)$.
\subsection{Theoretical predictions of the properties of interfacial water}
The description of the structure of the air-water interface by theoretical models is quite challenging. The dipole moment of the water molecule strongly depends on the environment. In bulk, the dipole moment of a water molecule is about 2.4D, while in the vapor phase becomes 1.8D. Rigid models, such as the SPC/E, TIP3P or TIP5P have a fixed dipole moment, irrespective of whether the water molecule is in bulk at the interface or in a gas phase. In fact, the surface tension of the vapor-water interface calculated with the most popular rigid water models\cite{Ismail2006,Vega2007} show significant disagreement (20\% or more) when compared with experimental results. Molecular dynamics ab-initio calculations\cite{Feng2004} report a slightly more expanded molecular area at the air-water interface, in qualitative agreement with the present study.

\subsection{Summary}
The objectives of the present study were twofold: first, to demonstrate that the structure factor of liquids can be accurately determined by GIXD in reflection mode, both to investigate the structure of the bulk or the interface, and second, to introduce a new method to compute the PDF from the structure factor in a way that allows to assess the intrinsic errors associated with the structure factor $S(q)$.

One advantage of measurements in reflection mode is that the raw intensities are very close to the actual $S(q)$ to within a factor, especially at large incident beam angles, as argued earlier\cite{Levy1966}. The major correction needed is the effective volume of scattering, which under suitable choices of slits could even be reduced to a trivial scaling factor independent of $q$. This should be contrasted with transmission measurements, where the measured intensity is dramatically different from the structure factor due to the effect of the container, and geometry (see Figure 4 in Ref. \onlinecite{Hura2000}).  A unique advantage of the X-ray in reflection-GIXD mode is that it can be applied to determine the structure at the vapor/liquid interface by adjusting the incident beam below the critical angle.  In this configuration the evanescent wave scatters mainly from the  topmost layers at the surface, and provide valuable information on the restructuring at the gas/liquid interface.

Our experimental results show that the bulk PDF does not describe the surface scattering data correctly, suggesting a restructuring of water molecules at the vapor/water interface. In particular, our results show that the water molecules at the top most layers are more expanded.

In the present study, we have also introduced a method based on a non-linear-least-square refinement to obtain the PDF from the experimental $S(q)$. This method avoids the problems associated with computing the Fourier transform by extrapolations to $q$-values that are not measured, and provides the uncertainties associated with the PDF that are compatible with the measured experimental data for $S(q)$.  We hope that our findings will initiate future experimental and theoretical studies of liquids surfaces in general.
\begin{appendix}
\section{Effective scattering volume, $V_{eff}$}
\begin{figure}[htl]
\includegraphics[width=2.4 in]{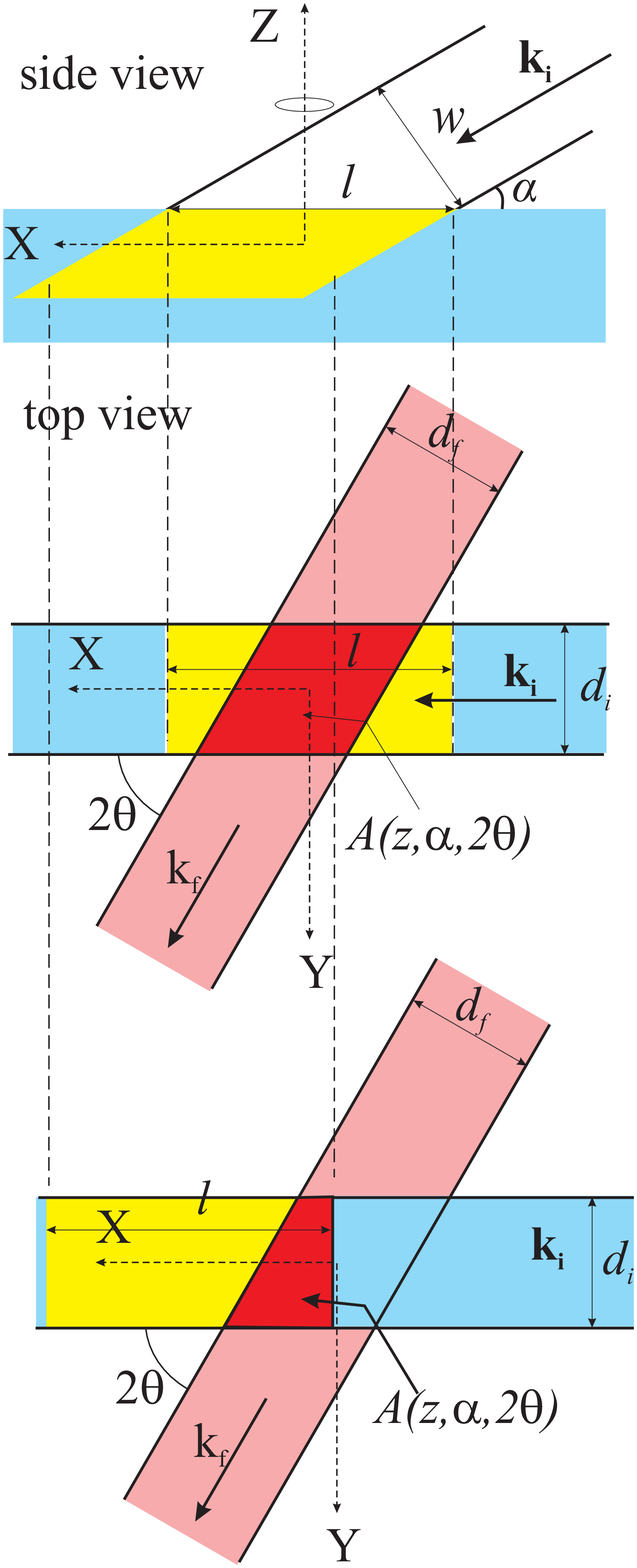}
\caption{\label{setup2}(A) Illustration of a side view of the beam footprint. As the beam penetrates the bulk the center of the foot print is shifted along X. (B) Top view of the beam footprint ($z=0$) cross section with the footprint of an out-going beam at angle $2\theta$. (C) Same as (B) but for a footprint of the incident beam at finite $z$ value.}
\end{figure}

The volume of scattering is defined as that region of the illuminated sample whose scattered rays are actually detected.  This volume depends on the angles of the incident and scattered beam with respect to the surface, the apertures of the detector, and the attenuation length through the sample (i.e., X-ray energy).
In our setup, the cross section of the incident beam is rectangular defined by two sets of slits with vertical and horizontal opening $w_i$ and  $d_i$, respectively.  Typically, $w_i \approx 0.01 -0.1 $ mm, and $d_i \approx 1 - 2$ mm.  The area $A_i$ of the incident beam footprint on the liquid surface varies with the angle of incidence $\alpha$
\begin{equation}
A_i(z; \alpha) = \frac{w_id_i}{\sin\alpha}=d_il,
\end{equation}
as illustrated in Fig.\ \ref{setup2}.
Although the area $A_i$ is preserved as the beam penetrates the bulk of the liquid, the center of the illuminated rectangle shifts away from the sample axis of rotation, changing the effective volume of scattering.  The outgoing beam slits, typically $w_f \approx 2 $ mm and $d_f \approx d_i$ vertically and horizontally, respectively, form a footprint that is longer than that formed by the incident beam.
The effective volume is an integral of the overlap area between the incoming and outgoing footprint at each $z$ value weighted by the attenuation length of the incident and scattered beams, as follows
\begin{equation}
V_{eff} = \int_0^{nL}A(z,\alpha,2\theta){\rm e}^{-z/L}{\rm d}z
\label{Veff}
\end{equation}
where $L \equiv (D(\alpha)D(\beta)/[D(\alpha)+D(\beta)]$ is the effective attenuation length the beam, and $n \approx 8$ ensures the convergence of the numerically calculated integral. The attenuation length into the bulk at an angle of incidence $\alpha$ is given by
\begin{equation}
D(\alpha)=1/\mathop{\rm Im}(k_z)
\label{Im_kz}
\end{equation}
where,
\begin{equation}
 k_z \left( z \right) = k_0 \sqrt {\sin ^2  \alpha   - 2\delta  - i2\gamma},
 \end{equation}
 and
\begin{equation}
\delta  = \frac{1}{{2\pi }}\sum\limits_j {N_j } r_0 \lambda ^2 f_j^{'}\left( {\lambda } \right);  \hspace{0.5cm}
\gamma  = \frac{1}{{2\pi }}\sum\limits_j {N_j} r_0\lambda ^2 f_j^{''} \left( {\lambda } \right)
\end{equation}
where $N_j$ is the number density of atom type $j$ with  form factor $f_j^{'}$ and absorption factor $f_j^{''}$.
\begin{figure}[htl]
\includegraphics[width=2.8 in]{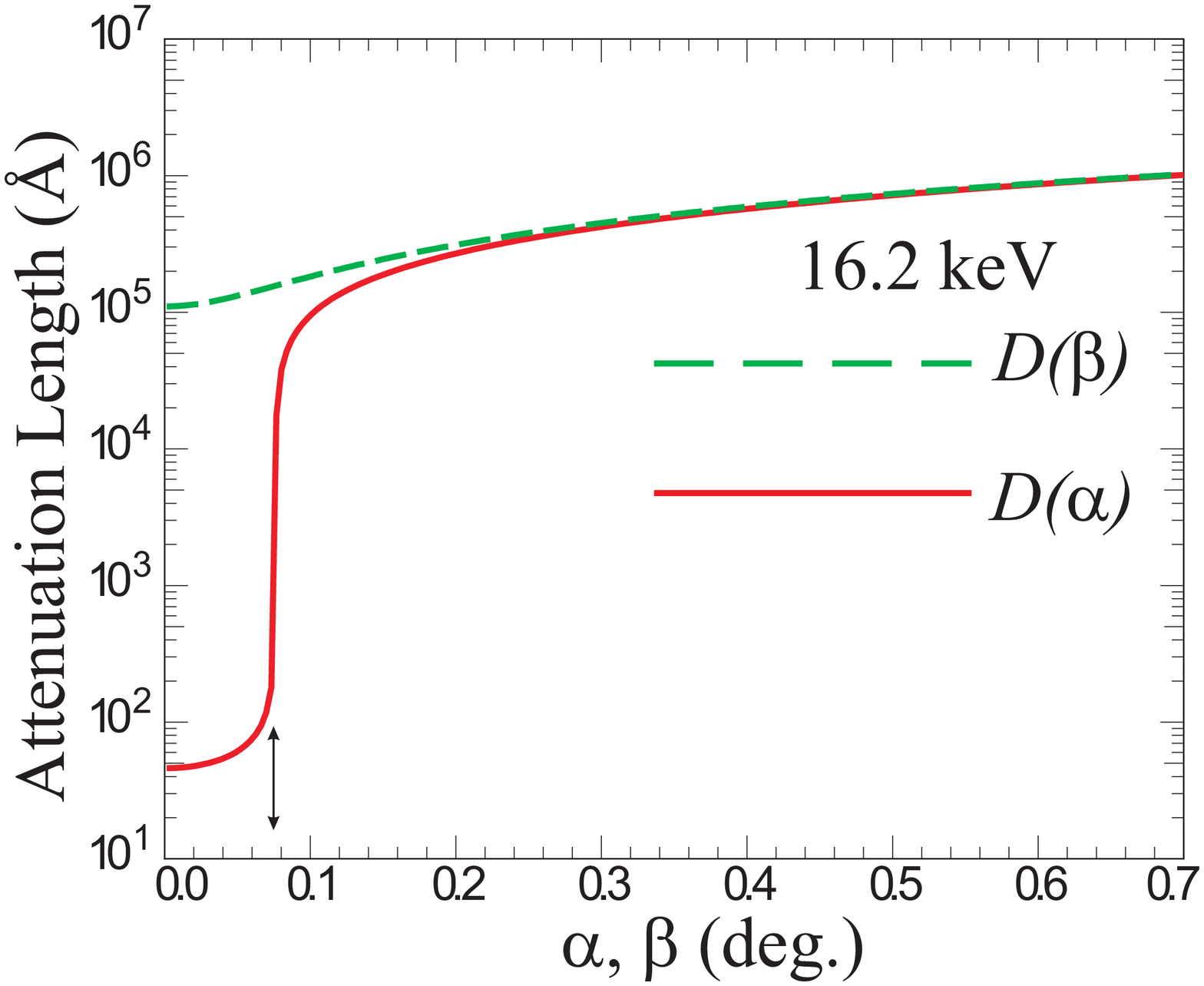}
\caption{\label{depth} Attenuation length for an external incident beam (solid line) and for an internal incident beam (dashed line) at the vapor/water interface, calculated by Eq.\ \ref{Im_kz} for a 16.2 keV X-ray beam. The two curves converge at angles larger than the critical angle for total reflection. (Arrow indicates the location of the critical angle)}
\end{figure}

Figure\ \ref{depth} shows the attenuation length as a function of the incident angle of a 16.2 keV X-ray beam propagating from the gas phase onto the liquid surface.  Below a critical incident angle for total external  reflection $\alpha_c$ the beam at the surface is evanescent; it penetrates to a finite depth ($D(\alpha)\leq 80$ {\AA}) into the bulk and emerges almost totally without transmission (a small fraction, less than 0.1{\%}, depending on the surface- roughness, scatters as diffuse scattering).  Above the critical angle, the beam is mostly transmitted but it is also attenuated by the absorption coefficient ($\gamma$) of the liquid.  The beam that scatters from the bulk at angle $\beta$ emerges through the surface with no total reflection, as implied in Figure \ref{depth}.  The two curves, $D(\alpha)$ and $D(\beta)$ converge at large angles.

The effective scattering volume is calculated from Eq.\ (\ref{Veff}) by determining the effective scattering area $A(z, \alpha,2\theta)$ at depth z. This area is polygonal in shape (see Fig.\ \ref{setup2}) and is determined from the geometrical constraints

\begin{equation}
-\frac{l}{2}-\frac{z}{\tan{\alpha}}\leq x \leq \frac{l}{2}-\frac{z}{\tan{\alpha}},
\end{equation}
which account for the shift of the footprint center as the beam penetrates into the bulk in the $z$ direction,
\begin{equation}
-\frac{d_i}{2}\leq y \leq \frac{d_i}{2}
\end{equation}
and
\begin{equation}
-\frac{d_f}{2\sin(2\theta)}+\frac{y}{\tan(2\theta)}\leq x \leq \frac{d_f}{2\sin(2\theta)}+\frac{y}{\tan(2\theta)}.
\label{Boundaries}
\end{equation}
\begin{figure}[thl]
\includegraphics[width=2.6 in]{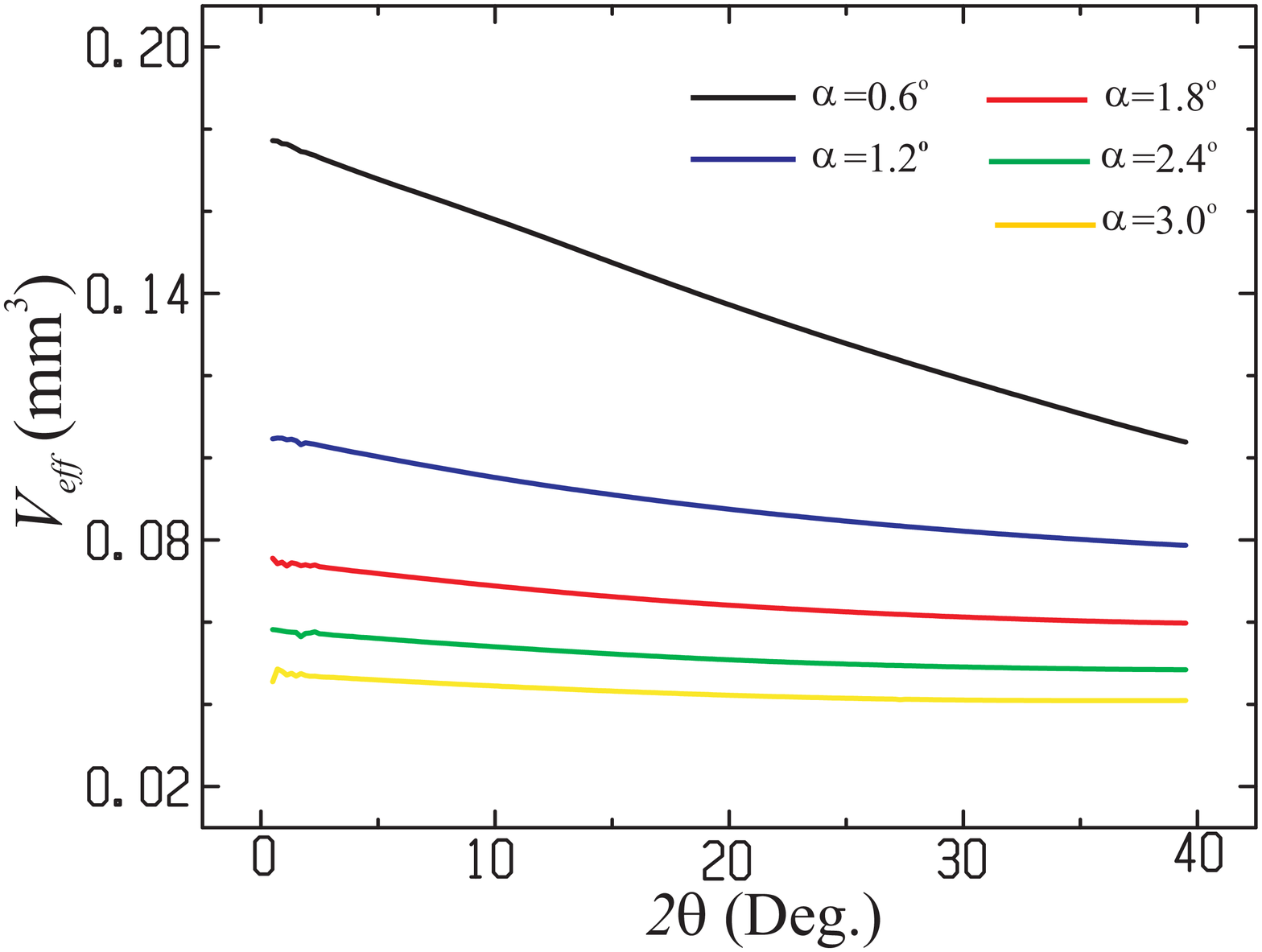}
\caption{\label{eff_volume} Effective volume of scattering using Eq.\ (\ref{Veff}) at various angles of incident beam $\alpha$ and $\beta=0.27^o$.   The calculation is for a 16.2 keV beam, and slit parameters $d_i= 1.8, d_f=2$ mm, $w=0.05$ mm.}
\end{figure}

Figure\ \ref{eff_volume} shows the effective scattering volume as a function of the angle $2\theta$ for different values of incident beam angle $\alpha$.
When $\alpha$ is relatively large, the footprint becomes smaller than the detector aperture and as a result, almost all of the illuminated area is detected, that is, the effective volume of scattering is almost insensitive to the scattering angle $2\theta$. As $\alpha$ becomes smaller, the overlap of incoming and outgoing footprints is more complex, resulting in a stronger variation of $V_{eff}$ on $2\theta$, as intuitively clear from Fig.\ \ref{setup2}.

\section{Calculation of scattering from half-filled space surface assuming bulk PDF \label{App-2}}
According to the diffuse scattering theory \cite{Sinha1988,Fukuto2006},
\begin{eqnarray}
I_{CW}(\alpha,\beta,2\theta)=aPA(0,\alpha,2\theta)\nonumber \\
\times\frac{(2q_{xy}+\Delta q_{xy})^{\eta}-(2q_{xy}-\Delta q_{xy})^{\eta}}{2^{\eta}} & \hfil\nonumber \\
\propto\frac{P}{q_{xy}^{2}}\;(q_{xy}\gg\Delta q_{xy};\eta\rightarrow0),\label{eq_sur2}\end{eqnarray}
 where $a$ is a scale factor, $P$ is the polarization factor, $A$
is the effective illumination area at $z=0$ (see Fig. 7), $q_{xy}=\sqrt{q_{x}^{2}+q_{y}^{2}}$,
$\Delta q_{xy}=k_{0}\cos\theta\Delta\theta$, $\eta=\frac{k_{B}T}{2\pi\gamma}q_{z}^{2}$,
and $\Delta\theta$ is the acceptance angle of the detector.

\begin{figure}[!h]
\includegraphics[width=2.4in]{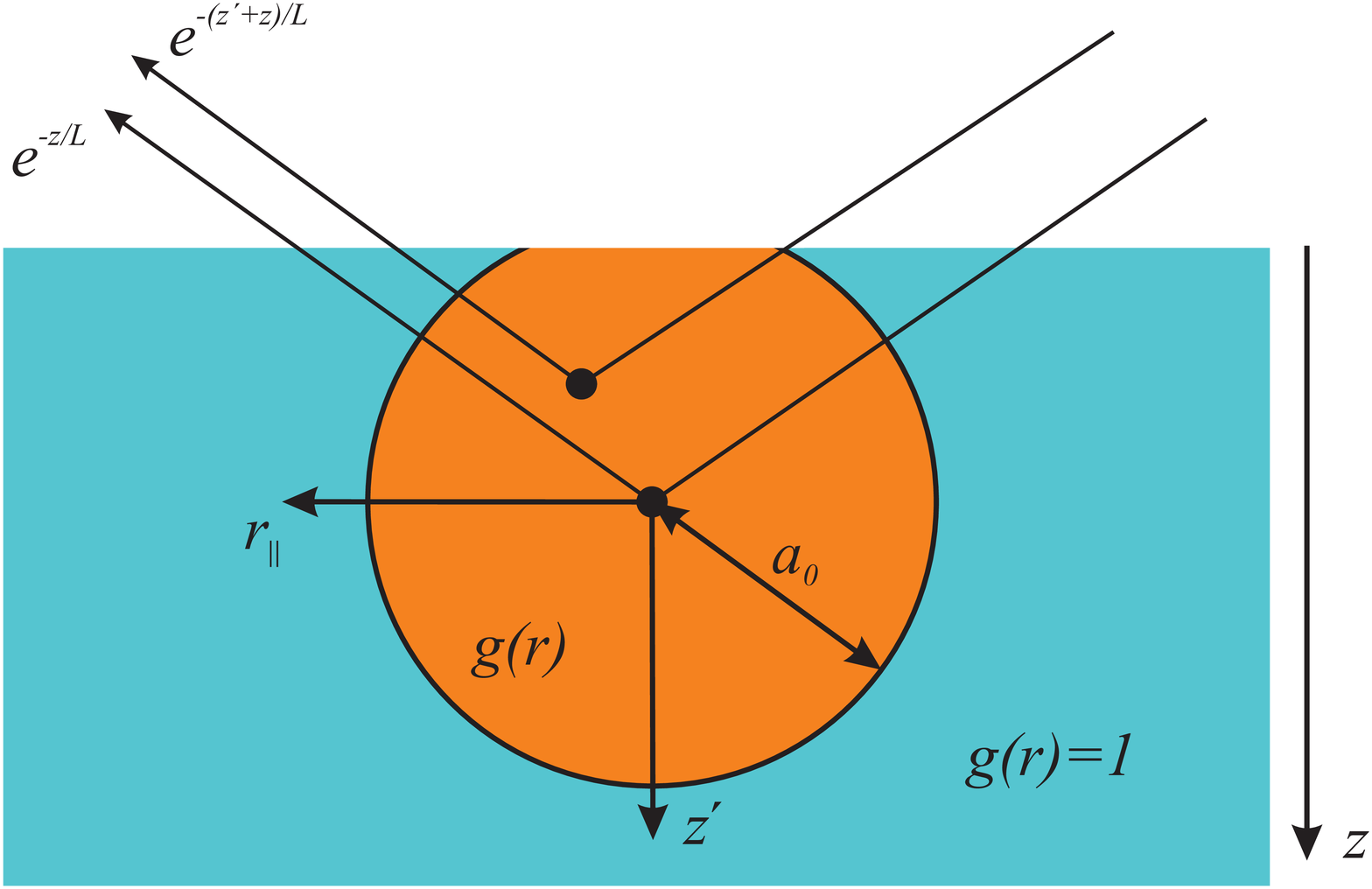}
\caption{\label{sur2} Illustrations for attenuation factors and the integration
range of Eq. \ \ref{eq_sur4}}
\end{figure}

By assuming $g(r)$ is the same everywhere including the region close
to the surface, the bulk scattering from a half-filled space with
a penetration depth $L$ (see Eq.\ (\ref{Veff})) can be written
as \begin{equation}
I_{b}(\alpha,\beta,2\theta)=bP\int_{0}^{nL}A(z,\alpha,2\theta)w(q,z)dz,\label{eq_sur3}\end{equation}
 where $b$ is a scale factor, $n=8$ as discussed in Eq. (\ref{Veff}),
and $w(q,z)$ is the scattering intensity from one water molecule
at a depth $z$ with respect to the surface given by \begin{eqnarray}
w(q,z)=e^{-z/L}<F^{2}>+<F>^{2}\int_{-z}^{a_{0}}\nonumber \\
\times e^{-(2z+z^{\prime})/2L}dz^{\prime}\nonumber \\
\times\int_{0}^{a_{0}}dr_{\parallel}\int_{0}^{2\pi}d\theta\rho r_{\parallel}(g(\sqrt{r_{\parallel}^{2}+z^{\prime2}})-1)e^{-iq_{z}z^{\prime}}\nonumber \\
e^{-iq_{xy}r_{\parallel}\cos\theta},\label{eq_sur4}\end{eqnarray}
 where $\rho$ is the number density of the water, $e^{-z/L}$ and
$e^{-(2z+z^{\prime})/2L}$ are the attenuation factors for the inelastic
and elastic scattering, respectively. To make the numerical integration
more efficient, we define a sphere of radius $a_{0}$ ($\sim12$ {\AA})
outside which $g(r)=1$ (Fig. \ \ref{sur2}) for $L\gg a_{0}$. One
should note that $w(q,z)=e^{-z/L}S(q)$ (see Eq. (\ref{SQ})) for
$z>a_{0}$. For large incident angles, $L\gg a_{0}$, Eq. \ (\ref{eq_sur3})
can be simplified to \begin{equation}
I_{b}\simeq bS(q)P\int_{a_{0}}^{nL}e^{-z/L}A(z,\alpha,2\theta)dz\simeq bS(q)PV_{eff},\label{eq_sur5}\end{equation}
 which is practically the same as Eq.\ (\ref{I_to_SQ}). For small
incident angles, $q_{z}\sim0$ so, to simplify the integrals in Eq.
\ (\ref{eq_sur4}) we assume $q_{z}=0$ yielding \begin{eqnarray}
w(q,z)\simeq e^{-z/L}<F^{2}>+<F>^{2}\nonumber \\
\times\int_{-z}^{a_{0}}e^{-(2z+z^{\prime})/2L}dz^{\prime}\nonumber \\
\times\int_{0}^{a_{0}}dr_{\parallel}2\pi\rho r_{\parallel}(g(\sqrt{r_{\parallel}^{2}+z^{\prime2}})-1)J_{0}(q_{xy}r_{\parallel}),\label{eq_suf6}\end{eqnarray}
 where $J_{0}$ is the zero order Bessel function.
\section{Application of the NLSF method to other $S(q)$ measurements}
\label{App-3}
\begin{figure}[thl]
\includegraphics[width=2.4 in]{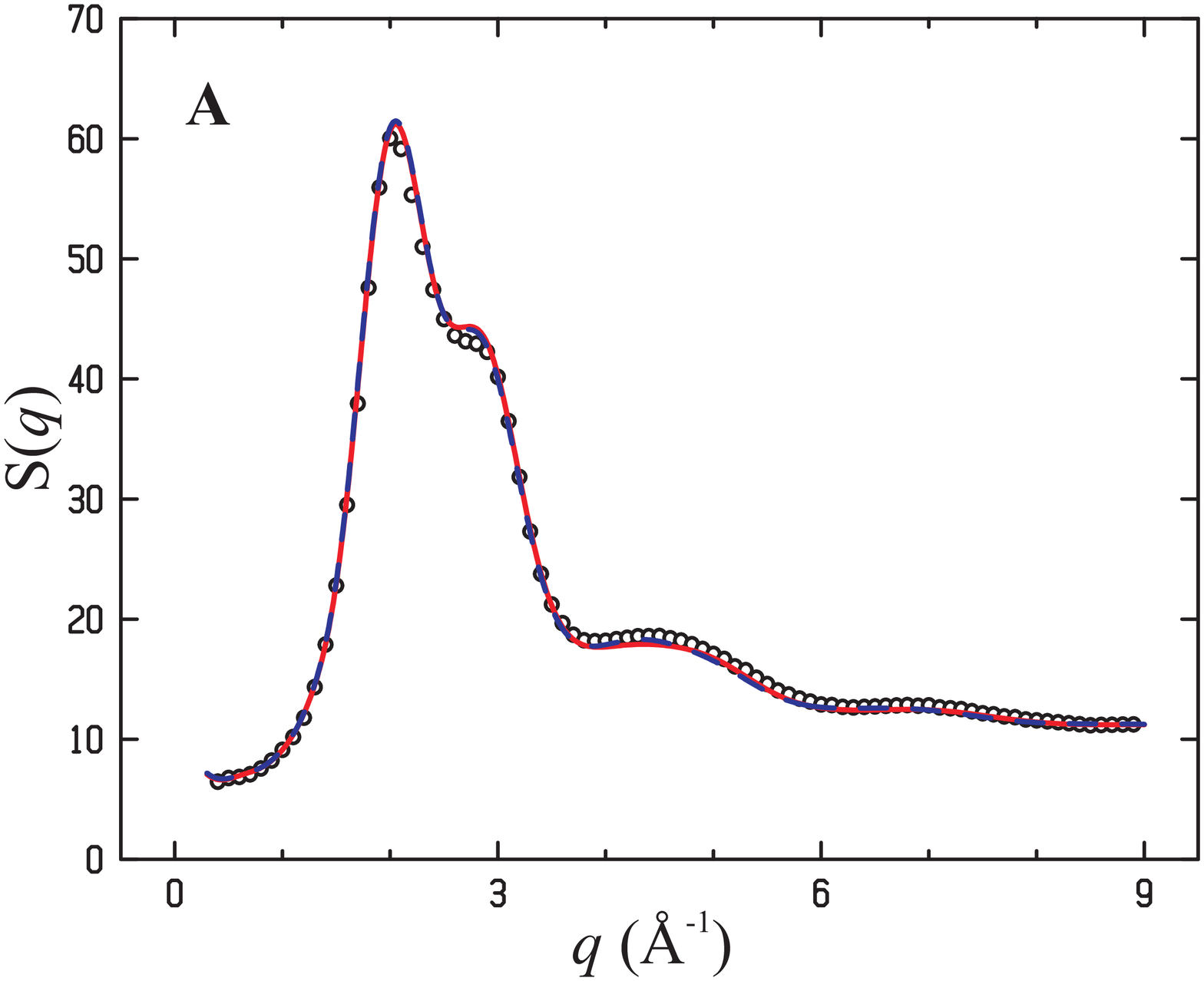}
\includegraphics[width=2.4 in]{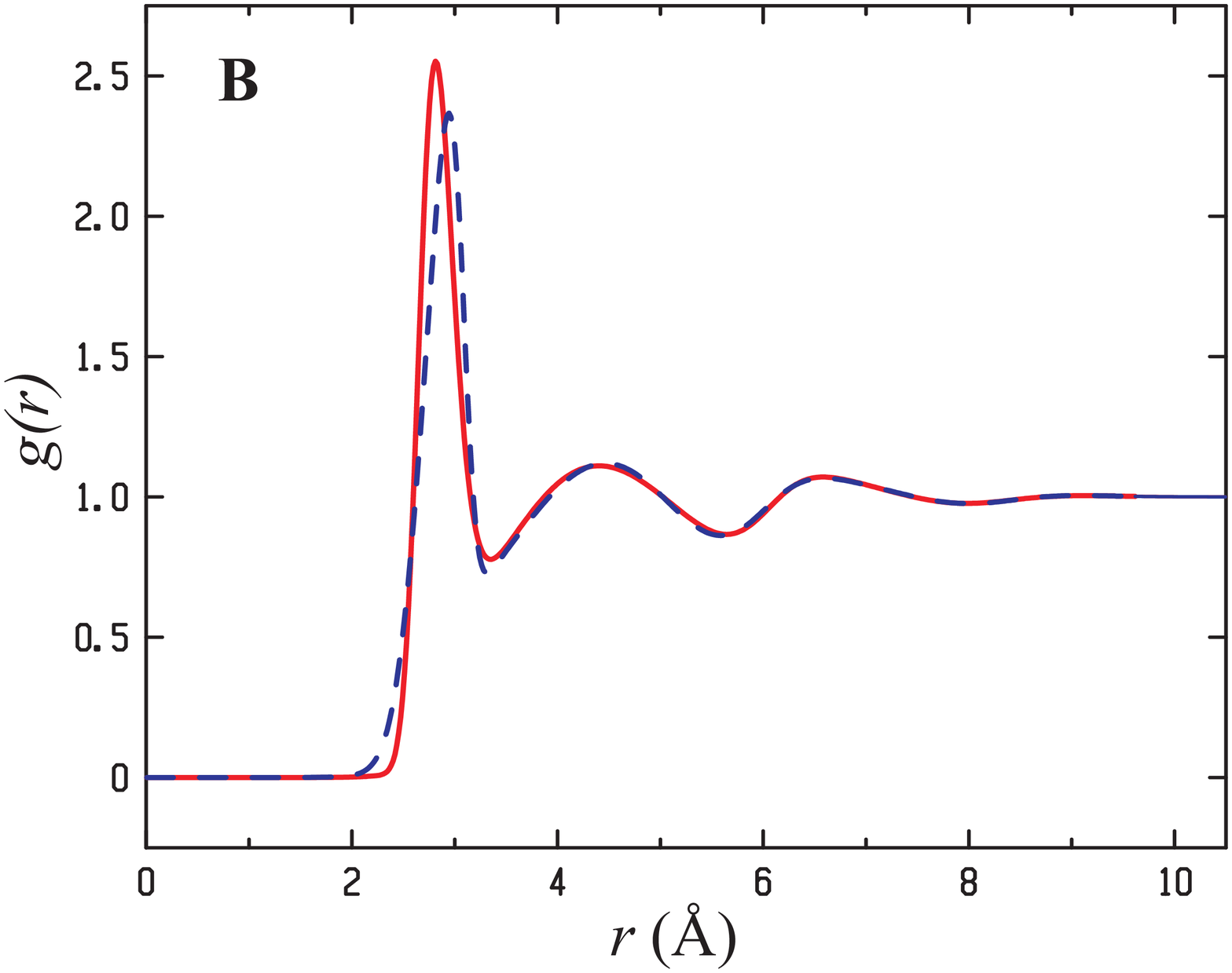}
\caption{\label{Comp_hura} (A) $S(q)$ data from Ref.\ \cite{Hura2000} (circles).  Solid- and dashed- lines are the best fits using the method described in Section\ \ref{sec-method}.   B) Two extracted PDFs from (A) that fit the data equally well. Despite the higher range of the measured $q$-values, possible uncertainty in height, width, and position of the first peak in $g(r)$ is evident.}
\end{figure}
Figure\ \ref{Comp_hura}(A) shows two slightly different fits (solid and dashed lines) to the $S(q)$ data in Ref.\ \onlinecite{Hura2000} using the method described in Section\ \ref{sec-method} to extract the PDF.  Although the two fits are practically of the same quality they produce different shapes $g(r)$'s in particular near the first peak, as shown in Fig.\ \ref{Comp_hura}(B).  We show this to demonstrate that even with data measured to larger $q$ values than in the present study, some ambiguity in the evaluation of the first molecular shell around a water molecule is still present.

\end{appendix}
\begin{acknowledgments}
We thank C. Lorentz for helpful insights on SPC/E and TIP3P models.  The MUCAT sector at the APS is supported by the U.S. DOE Basic Energy Sciences, Office of Science, through Ames Laboratory under contract No. W-7405-Eng-82. Use of the Advanced Photon Source is supported by the U.S. DOE, Basic Energy Sciences, Office of Science, under Contract No. W-31-109-Eng-38.  AT is supported by  NSF grant DMR-0426597 and partially by DOE-BES through the Ames lab under contract no. DE-AC02-07CH11358.
\end{acknowledgments}


\begin{references}
\bibitem{comment1} This statement is not a unique characteristic of the reflection mode.  The variation in intensities are also expected by changing the incident angle with respect to the normal to the windows of the container of the liquid, but to our knowledge, such study has not been done.
\end{references}
\end{document}